\documentclass[structabstract,openany]{aa}  

\usepackage{amsmath}
\usepackage{color}
\usepackage{footnote}
\makesavenoteenv{tabular}

\usepackage{natbib}
\bibpunct{(}{)}{;}{a}{}{,} 

\usepackage{marvosym}
\usepackage{hyperref}
\hypersetup{
bookmarks=true,
pdftoolbar=true,
pdfmenubar=true,
colorlinks,breaklinks,
linkcolor=blue,urlcolor=blue,
anchorcolor=blue,citecolor=blue}

\usepackage[varg]{txfonts}
\begin{document}

\newcommand{\kms}{km s$^{-1}$}

\title{The chemistry of ions in the Orion Bar I. - CH$^+$, SH$^+$, and CF$^+$
\thanks{{\em Herschel} is an ESA space observatory with science instruments provided by European-led Principal Investigator consortia and with important participation from NASA.}}
\subtitle{The effect of high electron density and vibrationally excited H$_2$ in a warm PDR surface}
\authorrunning{Z. Nagy et al.}

\author{Z. Nagy\inst{1,2},
        F.F.S. Van der Tak\inst{2,1},          
        V. Ossenkopf\inst{3},
		M. Gerin\inst{4},
		F. Le Petit\inst{5},
		J. Le Bourlot\inst{5},
        J. H. Black\inst{6}, 
        J. R. Goicoechea\inst{7},
		C. Joblin\inst{8,9},	
		M. R\"{o}llig\inst{3},
		\and
		E. A. Bergin\inst{10}		  
        }

\institute{
Kapteyn Astronomical Institute, PO Box 800, 9700 AV, Groningen, The Netherlands \\
\email{nagy@astro.rug.nl} 
\and
SRON Netherlands Institute for Space Research, PO Box 800, 9700 AV, Groningen, The Netherlands 
\and
I. Physikalisches Institut der Universit\"{a}t zu K\"{o}ln, Z\"ulpicher Stra{\ss}e 77, 50937 K\"oln, Germany 
\and 
LERMA, UMR 8112 du CNRS, Observatoire de Paris, \'Ecole Normale Sup\'erieure, France
\and
Observatoire de Paris, LUTH and Universit\'e Denis Diderot, Place J. Janssen, 92190 Meudon, France
\and     
Chalmers University of Technology, Department of Earth and Space Sciences, Onsala Space Observatory, 43992 Onsala, Sweden
\and
Centro de Astrobiolog\'ia, CSIC-INTA, 28850 Madrid, Spain
\and
Universit\'e de Toulouse, UPS-OMP, IRAP, Toulouse, France
\and
CNRS, IRAP, 9 Av. colonel Roche, BP 44346, 31028 Toulouse Cedex 4, France
\and
Department of Astronomy, The University of Michigan, 500 Church Street, Ann Arbor, MI 48109-1042, USA
}

\date{Received October 8, 2012; accepted December 12, 2012}

 
  \abstract
   {The abundances of interstellar CH$^+$ and SH$^+$ are not well understood as their most likely formation channels are highly endothermic. 
   Several mechanisms have been proposed to overcome the high activation barriers, including shocks, turbulence, and H$_2$ vibrational excitation.}
   {Using data from the Herschel Space Observatory, we studied the formation of ions, in particular CH$^+$ and SH$^+$ in a typical high UV-illumination warm and dense photon-dominated region (PDR), the Orion Bar.} 
   {The HIFI instrument on board Herschel provides velocity-resolved line profiles of CH$^+$ 1-0 and 2-1 and three hyperfine transitions of SH$^+$ $1_2-0_1$. The PACS instrument provides information on the excitation and spatial distribution of CH$^+$ by extending the observed CH$^+$ transitions up to $J=6-5$. We compared the observed line intensities to the predictions of radiative transfer and PDR codes.}
   {All CH$^+$, SH$^+$, and CF$^+$ lines analyzed in this paper are seen in emission. The widths of the CH$^+$ 2-1 and 1-0 transitions are of $\sim$5 \kms, significantly broader than the typical width of dense gas tracers in the Orion Bar ($\sim$2-3 \kms) and are comparable to the width of species that trace the interclump medium such as C$^+$ and HF. The detected SH$^+$ transitions are narrower compared to CH$^+$ and have line widths of $\sim$3 \kms, indicating that SH$^+$ emission mainly originates in denser condensations. Non-LTE radiative transfer models show that electron collisions affect the excitation of CH$^+$ and SH$^+$ and that reactive collisions need to be taken into account to calculate the excitation of CH$^+$. Comparison to PDR models shows that CH$^+$ and SH$^+$ are tracers of the warm surface region (A$_{\rm{V}}<$1.5) of the PDR with temperatures between 500 and 1000 K. We have also detected the 5-4 transition of CF$^+$ at a width of $\sim$1.9 \kms, consistent with the width of dense gas tracers. The intensity of the CF$^+$ 5-4 transition is consistent with previous observations of lower$-J$ transitions toward the Orion Bar.}
   {An analytic approximation and a numerical comparison to PDR models indicate that the internal vibrational energy of H$_2$ can explain the formation of CH$^+$ for typical physical conditions in the Orion Bar near the ionization front. The formation of SH$^+$ is also likely to be explained by H$_2$ vibrational excitation.
The abundance ratios of CH$^+$ and SH$^+$ trace the destruction paths of these ions, and indirectly, the ratios of H, H$_2$, and electron abundances as a function of depth into the cloud.}


   \maketitle
%

\section{Introduction} 

The methylidyne cation CH$^+$ was one of the first molecules to be detected in the interstellar medium \citep{douglasherzberg1941}. Early studies of CH$^+$ found its abundance to be consistently larger than the predictions of steady-state chemical models in quiescent molecular clouds (e.g. \citealp{vandishoeckblack1986}). One of the possible formation routes is the endothermic reaction $\mathrm{C^+ + H_2 + ~0.41~eV}$ $\mathrm{\rightarrow CH^+ + H}$.
To reproduce the observed CH$^+$ abundances, several mechanisms have been proposed to overcome the high activation barrier of the formation reaction. For low-density diffuse interstellar clouds, C-shocks \citep{pineaudesforets1986} and turbulent dissipation \citep{godard2009} have been proposed and confirmed by \citet{falgarone2010a,falgarone2010b} and \citet{godard2012}. 
Alternatively, in denser regions with strong far-ultraviolet (FUV) radiation fields, the internal energy available in the vibrationally excited H$_2$ molecules has been proposed to help overcome the large activation barrier (\citealp{sternberg1995}, \citealp{agundez2010}).

Sulfanylium (SH$^+$) has a similar chemistry to CH$^+$, having a formation route via S$^+$ and H$_2$; however, this reaction is twice as endothermic as the CH$^+$ formation reaction. After non-detections of SH$^+$ in the UV domain in the spectra of nearby stars (\citealp{millarhobbs1988}, \citealp{magnanisalzer1989,magnanisalzer1991}), the 526 GHz $N_J=1_2-0_1$ transition of SH$^+$ has been detected in emission using Herschel toward the high-mass protostar W3 IRS 5 \citep{benz2010}. The 526 GHz transition has also been detected in absorption in the diffuse interstellar medium towards various distant star-forming regions \citep{godard2012}. The 683 GHz transition of SH$^+$ has been detected in absorption towards Sgr B2(M) from the ground with the Carbon Heterodyne Array of the MPIfR (CHAMP+) receiver of the Atacama Pathfinder EXperiment 12 m telescope (APEX) \citep{menten2011}.

In this paper, we study the formation and excitation of CH$^+$ and SH$^+$ in a prototypical photon-dominated region (PDR), the Orion Bar. 
The Orion Bar is located at a distance of 414 pc \citep{menten2007}. Its stratified structure has been the subject of many previous studies (such as \citealp{vanderwiel2009} and references therein). The mean density of the Orion Bar is about $\gtrsim$10$^5$ cm$^{-3}$, the mean molecular gas temperature 85 K \citep{hogerheijde1995}, and the impinging radiation field is $(1-4) \times 10^4$ in Draine units (Draine field: $\chi = 2.7\times10^{-3}\rm{erg}~\rm{s}^{-1}~ \rm{cm}^{-2}$ for the energy range 6 $<$ h$\nu$ $<$ 13.6 eV; \citealp{draine1978}). 
Most of the low-$J$ molecular line emission originates in an interclump medium with a density between a few 10$^4$ and 2$\times$10$^5$ cm$^{-3}$ \citep{simon1997}. 
High-density tracers such as HCN and H$^{13}$CN originate in dense clumps, as confirmed by interferometric observations \citep{youngowl2000}. The density of the clumps is in the range between 1.5$\times$10$^6$ and 6$\times$10$^6$ cm$^{-3}$ \citep{lis2003}.
Apart from the large clumps detected in H$^{13}$CN deep inside the Bar, small, warm ($T_\mathrm{kin}\sim160-220$ K), and dense ($n_{\rm{H}}\sim10^{6-7}$ cm$^{-3}$) condensations have been suggested to explain the excited OH emission at the PDR surface \citep{goicoechea2011}.

This paper aims to characterize the medium where ions such as CH$^+$ and SH$^+$ form and to distinguish between the mechanisms that can overcome the high activation barriers of the formation reaction in a warm and dense PDR. We also report the detection of the CF$^+$ 5-4 transition. We will address another ion, OH$^+$ in a separate paper (Van der Tak \& Nagy et al., in preparation).

\section{Observations and Data reduction}
\label{obs}

The CO$^+$ peak ($\alpha_\mathrm{J2000}=\rm{05^h35^m20.6^s}$, $\delta_\mathrm{J2000}=-05^\circ 25'14''$) in the Orion Bar \citep{storzer1995} has been observed as part of the Herschel observations of EXtra-Ordinary Sources (HEXOS) guaranteed-time key program \citep{bergin2010} for the Heterodyne Instrument for the Far-Infrared (HIFI, \citealp{degraauw2010}) of the Herschel Space Observatory \citep{pilbratt2010} in every HIFI band as a spectral scan.
The data were reduced using the \textit{Herschel} Interactive Processing Environment (HIPE, \citealp{ott2010}) pipeline version 6.0. The velocity calibration of HIFI data is accurate to $\sim$0.5 kms$^{-1}$ or better. The sideband deconvolution was done using the \textit{doDeconvolution} task in HIPE. In this paper we use the HIFI bands 1a, 1b, 2a and 3a from the HEXOS spectral line survey. These observations were carried out in March and April 2011 in load chop mode with a redundancy of 4 and had total integration times of 2.4 h, 2.2 h, 3.1 h, and 1.3 h, respectively. The Wide-Band Spectrometer (WBS) backend was used which covers 4 GHz bandwidth in four 1140 MHz subbands at 1.1 MHz resolution.

In addition to HIFI spectral scans observed as a part of the HEXOS key program, the CH$^+$ 2-1 transition was observed as a deep integration in a spectral scan in band 6b, with a total integration time of 11.7 h and a redundancy 4 in dual-beam-switch (DBS) mode. Both WBS and the High Resolution Spectrometer (HRS) backends were used. It was reduced using HIPE pipeline version 8.0.

Besides the HIFI data, we used observations of the CH$^+$ 3-2, 4-3, 5-4, and 6-5 transitions from the Photodetector Array Camera and Spectrometer (PACS, \citealp{poglitsch2010}) onboard Herschel.  
The PACS observations were carried out in September 2010 and consist of two spectral scans in Range Spectroscopy mode with 5 range repetitions each (Joblin et al. 2012, in preparation). The PACS spectrometer provides 25 spectra over a 47$''\times47''$ field-of-view resolved in 5$\times$5 spatial pixels (``spaxels''), each with a size of $\sim$9.4$''$ in the sky. The measured width of the spectrometer point spread function (PSF) is relatively constant at $\lambda\lesssim$100\,$\mu$m, but it significantly increases above the spaxel size for longer wavelengths. The resolving power varies between $\lambda$/$\Delta\lambda\sim$1000 (R1 grating order) and $\sim$5000 (B3A grating order). The central spaxel was centered at the same HIFI survey position. Observations were carried out in the ``chop-nodded'' mode with the largest chopper throw of 6~arcmin. The total integration time was 3.2\,h for the 1342204117 observation (B2B and R1) and 2.7\,h for the 1342204118 observation (B3A). PACS data were processed using HIPE 6.0.3.

\begin{center}
\begin{figure}[!h]
\centering
\includegraphics[width=8cm,angle=-90,trim=0cm 0cm 0 0,clip=true]{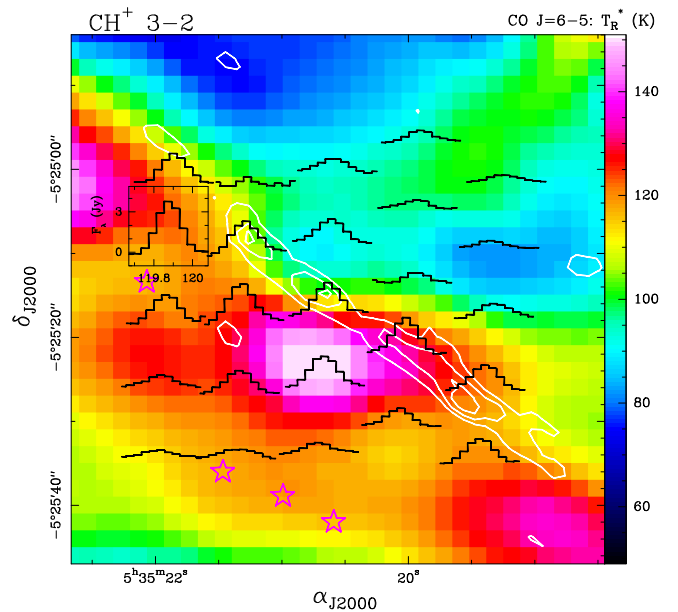}
\caption{PACS CH$^+$ 3-2 lines (black, centered on the CO$^+$ peak) overlaid on the distribution of the CO $J=6-5$ peak brightness temperature (color image) observed with the CSO telescope at $\sim$11$''$ resolution \citep{lis1998}. 
The PACS line intensity distributions are shown in units of Jy/spaxel as a function of wavelength in $\mu$m and are not velocity resolved. 
White contours show the brightest regions of H$^*_{2}~v=1-0$ S (1) emission \citep{walmsley2000}. 
Lower-intensity H$^*_2$ extended emission is present in the entire field \citep{vanderwerf1996}. Violet stars show the position of the H$^{13}$CN $J=1-0$ clumps deeper inside the Bar \citep{lis2003}.
}
\label{pacs_cso_maps_CHp32}
\end{figure}
\end{center}

Table \ref{table:line_sample} shows the spectroscopic and observational parameters of the transitions used in this paper. The rest frequencies are based on the Cologne Database for Molecular Spectroscopy (CDMS database, \citealp{muller2005}). In particular, the frequencies of the CH$^+$ 1-0 and 2-1 and $^{13}$CH$^+$ 1-0 transitions are based on \citet{muller2010}.

Table \ref{table:line_param} includes the observed line parameters for our line sample observed with HIFI, which have been corrected for main beam efficiencies based on \citet{roelfsema2012}. 
In the case of the detected HIFI transitions we use the average of H- and V-polarizations. The detected lines show similar line profiles in both polarizations. The intensity difference between the polarizations is 12\% for CH$^+$ $J=1-0$, 8\% for CH$^+$ $J=2-1$, $\sim$20\% for SH$^+$ $N_J=1_2-0_1$, and $\sim$30\% for CF$^+$ 5-4.
To compare the line intensities of the transitions detected with HIFI with different beam sizes, we convert all the observed line intensities to a common $\sim$47$''$ resolution. We derive conversion factors between the original beam sizes and $\sim$47$''$ based on the integrated intensity map of the HCN 4$-$3 transition from the James Clerk Maxwell Telescope (JCMT) Spectral Legacy Survey (\citealp{vanderwiel2009}). For this, we assume the intensity distribution of the HCN 4$-$3 transition to trace the spatial structure of the Orion Bar.
For the PACS data (Table \ref{table:line_param}), because of uncertainties in the PSF-correction, we use the mean value of the intensity measured at the central spaxel ($I_\mathrm{central}$) and that corresponding to the value integrated over 3$\times$3 spaxels around the center ($I_{3\times3}$). The corresponding error bars are the difference between $I_\mathrm{central}$ and $I_{3\times3}$. This method provides the correct agreement between Spectral and Photometric Imaging Receiver (SPIRE, \citealp{griffin2010}) and HIFI data for CO lines (Joblin et al. 2012, in preparation).
The large (40\%) error bar corresponding to the intensity of the 5-4 transition is caused by the presence of oscillations (fringes) in this range. 
Owing to a lower band intensity and a significant noise level, only $I_{3\times3}$ was measured for the intensity of the 6-5 transition; $I_{\rm{central}}$ was then estimated using a standard error value.

\begin{table*}[t]
\begin{savenotes}
\begin{minipage}[!h]{\linewidth}\centering
\caption{Spectroscopic and observational parameters of the transitions used in this paper.}
\label{table:line_sample}
\renewcommand{\footnoterule}{}
\begin{tabular}{lrrllll}
\hline
\hline
\footnotetext[1]{The size of one PACS spaxel.}
Transition& Frequency& $E_{\rm{up}}$&   A&          instrument/band& beam-size& $\eta_{\rm{mb}}$\\
          &     (MHz)&           (K)&   (s$^{-1}$)&     &    ($''$)&                 \\
\hline
$~~~$CH$^+$ 1$-$0&        835137.5&  40.1& 6.36$\times$10$^{-3}$& HIFI, band 3a& 26.5& 0.75\\
$~~~$CH$^+$ 2$-$1&       1669281.3& 120.2& 6.10$\times$10$^{-2}$& HIFI, band 6b& 15.0& 0.72\\

$~~~$CH$^+$ 3$-$2&       2501440.5& 240.2& 2.20$\times$10$^{-1}$& PACS& $9.4$\footnotemark[1]& \\
$~~~$CH$^+$ 4$-$3&       3330629.7& 400.1& 5.38$\times$10$^{-1}$& PACS& $9.4$\footnotemark[1]& \\
$~~~$CH$^+$ 5$-$4&       4155872.0& 599.5& 1.07& PACS& $9.4$\footnotemark[1]& \\
$~~~$CH$^+$ 6$-$5&       4976201.4& 838.3& 1.86& PACS& $9.4$\footnotemark[1]& \\
\hline
$^{13}$CH$^+$ 1$-$0& 830216.1&  39.9& 5.83$\times$10$^{-3}$& HIFI, band 3a& 26.5& 0.75\\          
\hline
$~~~$SH$^+$ $N_J=1_2-0_1$, $F=3/2-1/2$& 526038.7& 25.3& 7.99$\times$10$^{-4}$& HIFI, band 1a& 44.2& 0.76\\
$~~~$SH$^+$ $N_J=1_2-0_1$, $F=5/2-3/2$& 526047.9& 25.3& 9.59$\times$10$^{-4}$& HIFI, band 1a& 44.2& 0.76\\
$~~~$SH$^+$ $N_J=1_2-0_1$, $F=3/2-3/2$& 526124.9& 25.3& 1.60$\times$10$^{-4}$& HIFI, band 1a& 44.2& 0.76\\
$~~~$SH$^+$ $N_J=1_1-0_1$, $F=3/2-1/2$& 683336.1& 32.8& 2.90$\times$10$^{-4}$& HIFI, band 2a& 33.2& 0.75\\
$~~~$SH$^+$ $N_J=1_1-0_1$, $F=1/2-1/2$& 683362.0& 32.8& 1.16$\times$10$^{-3}$& HIFI, band 2a& 33.2& 0.75\\
$~~~$SH$^+$ $N_J=1_1-0_1$, $F=3/2-3/2$& 683422.3& 32.8& 1.45$\times$10$^{-3}$& HIFI, band 2a& 33.2& 0.75\\
$~~~$SH$^+$ $N_J=1_1-0_1$, $F=1/2-3/2$& 683448.2& 32.8& 5.79$\times$10$^{-4}$& HIFI, band 2a& 33.2& 0.75\\
\hline
$~~~$CF$^+$ 5-4& 512846.5&  73.8& 8.21$\times$10$^{-4}$& HIFI, band 1a& 44.2& 0.76\\
$~~~$CF$^+$ 6-5& 615365.6& 103.4& 1.44$\times$10$^{-3}$& HIFI, band 1b& 44.2& 0.76\\
\hline
$^{13}$CF$^+$ 5-4& 488664.3& 70.0& 7.10$\times$10$^{-4}$& HIFI, band 1a& 44.2& 0.76\\
\hline
\end{tabular}
\end{minipage}
\end{savenotes}
\end{table*}

\section{Results}

\subsection{The detected CH$^+$, SH$^+$, and CF$^+$ transitions}
 
\begin{figure}[!h]
\centering 
\includegraphics[width=8.5 cm,trim=0cm 0cm 0cm 0cm,clip=true]{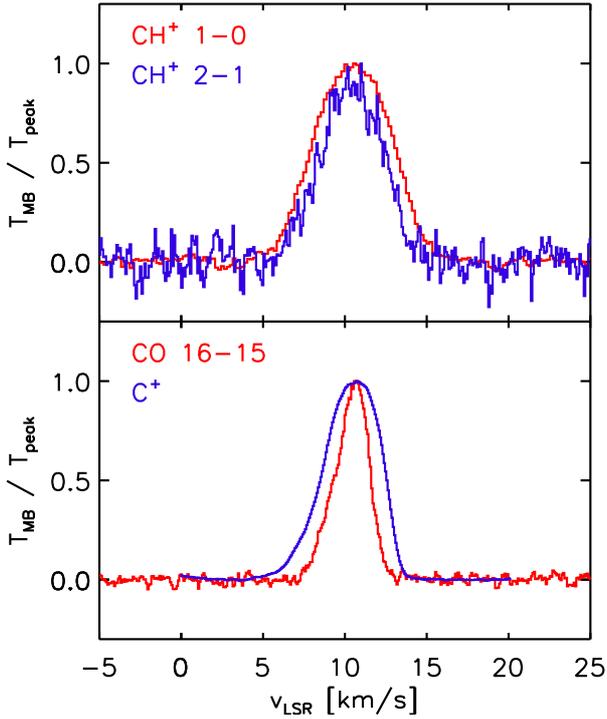} 
\caption{
\emph{Top panel:} Line profiles of CH$^+$ 2$-$1 and 1$-$0 transitions corresponding to the average of H- and V-polarizations observed with Herschel/HIFI toward the CO$^+$ peak in the Orion Bar.
\emph{Bottom panel:} Line profiles of C$^+$ and CO 16-15, for comparison, observed with Herschel/HIFI toward the CO$^+$ peak in the Orion Bar.}
\label{line_profiles_chplus}
\end{figure}

Figure \ref{line_profiles_chplus} shows the velocity-resolved CH$^+$ transitions that were detected with significantly broad lines ($\sim$5 \kms) compared to the line width of dense gas tracers in the Orion Bar ($\sim$2-3 \kms, \citealp{hogerheijde1995} and an example of the CO 16-15 transition in Fig. \ref{line_profiles_chplus}). The line width of the CH$^+$ 2-1 and 1-0 transitions is not only significantly larger than the width of dense gas tracers, but also of species that trace a similar region to CH$^+$, such as CO$^+$ (\citealp{storzer1995}, \citealp{fuente2003}). The 5 km s$^{-1}$ is comparable to the width of tracers of the interclump medium, such as C$^+$ ($\Delta v \sim 3.8$ km s$^{-1}$, also shown in Fig. \ref{line_profiles_chplus}) and HF ($\Delta v \sim 4.9$ km s$^{-1}$, \citealp{vandertak2012}).

Other non velocity-resolved transitions from PACS, such as the $J$=3-2 transition (Fig. \ref{pacs_cso_maps_CHp32}), show extended emission detected over all PACS spaxels, decreasing with distance from the ionization front. 
The maximum CH$^+$ 3-2 emission is seen farther into the nearly edge-on PDR compared to the peak H$_2$ $v$=1-0 emission (Figure \ref{pacs_cso_maps_CHp32}). As CH$^+$ forms via a reaction between C$^+$ and H$_2^*$, we compare the CH$^+$ 3-2 intensity distribution to C$^+$ emission from Figure 4 in \citet{ossenkopf2012}. The C$^+$ peak matches the CH$^+$ peak within the HIFI beam. This indicates that CH$^+$ formation is limited by the C$^+$ abundance rather than by the H$_2$ excitation that is traced by the H$_2$ $v=1-0$ S(1) emission observed by \citet{walmsley2000}.

\begin{figure}[!h]
\centering
\includegraphics[width=8.5 cm,trim=0.8cm 0cm 0 0,clip=true]{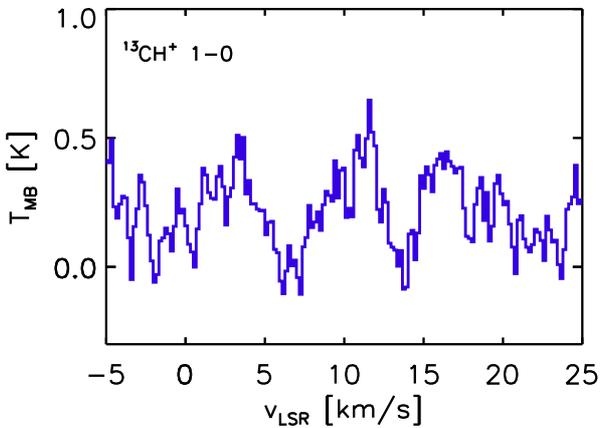} 	    
\caption{Line profile of $^{13}$CH$^+$ 1$-$0 transition observed with Herschel/HIFI in V-polarization toward the CO$^+$ peak in the Orion Bar.}
\label{line_profiles_c13hplus}
\end{figure}

Figure \ref{line_profiles_c13hplus} shows a tentative detection of the $^{13}$CH$^+$ $J=1-0$ in V-polarization. If it were a real detection, the observed $^{12}$CH$^+$/$^{13}$CH$^+$ line ratio of $\sim$40 would indicate an optical depth of the $^{12}$CH$^+$ line of $\sim$unity, assuming that the $^{13}$CH$^+$ emission is optically thin. However, this seems unlikely since a $^{12}$CH$^+$ optical depth of $\sim$1 at a temperature $>$100 K (see Sect. \ref{radex}) is inconsistent with the observed $^{12}$CH$^+$ line intensity. Deeper observations are needed to confirm the detection of $^{13}$CH$^+$.

\begin{figure}[!h]
\centering
\includegraphics[width=8.5 cm,trim=0.8cm 0.0cm 0.0 0.0,clip=true]{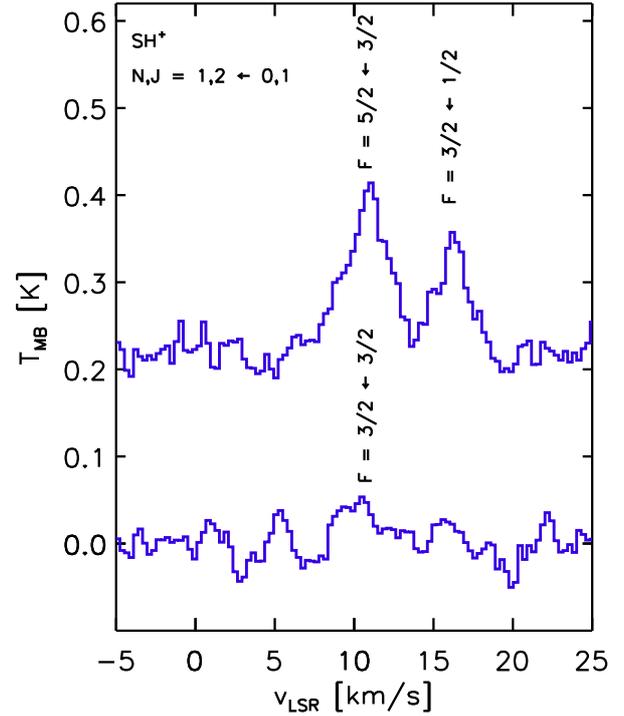} 	    
\caption{Line profiles of the three hyperfine transitions of SH$^+$ $N_J=1_2-0_1$ observed with Herschel/HIFI corresponding to the average of H and V polarizations toward the CO$^+$ peak in the Orion Bar. 
}
\label{line_profiles_shplus}
\end{figure}

Figure \ref{line_profiles_shplus} shows three hyperfine components of the $N_J=1_2-0_1$ transition of SH$^+$ ($F=3/2-1/2$, $F=5/2-3/2$, and $F=3/2-3/2$). The SH$^+$ $F$=3/2-3/2 transition is only detected in V-polarization. The line width of the detected SH$^+$ transitions (3.0 \kms) is narrower than those of CH$^+$ and are consistent with the width of dense gas tracers in the Orion Bar, suggesting that it does not originate in the same gas component as CH$^+$.

\begin{figure}[!h]
\centering
\includegraphics[width=8.5 cm,trim=0.8cm 0cm 0 0,clip=true]{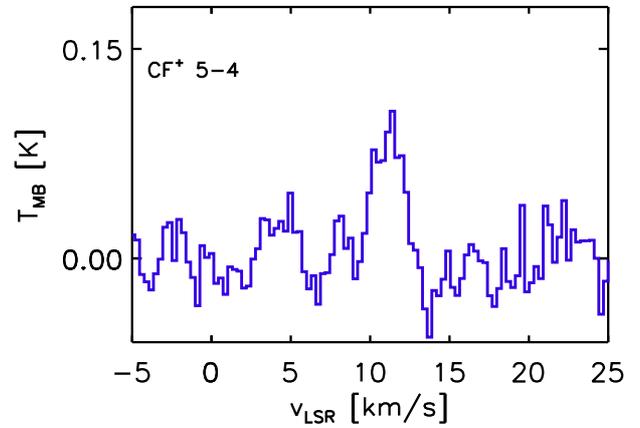} 	    
\caption{Line profile of CF$^+$ 5-4 transition corresponding to the average of H and V polarizations observed with Herschel/HIFI toward the CO$^+$ peak in the Orion Bar.}
\label{line_profiles_cfplus}
\end{figure}

\begin{figure}[!h]
\centering
\includegraphics[width=8.1 cm,trim=0.0cm 0cm 0 0,clip=true]{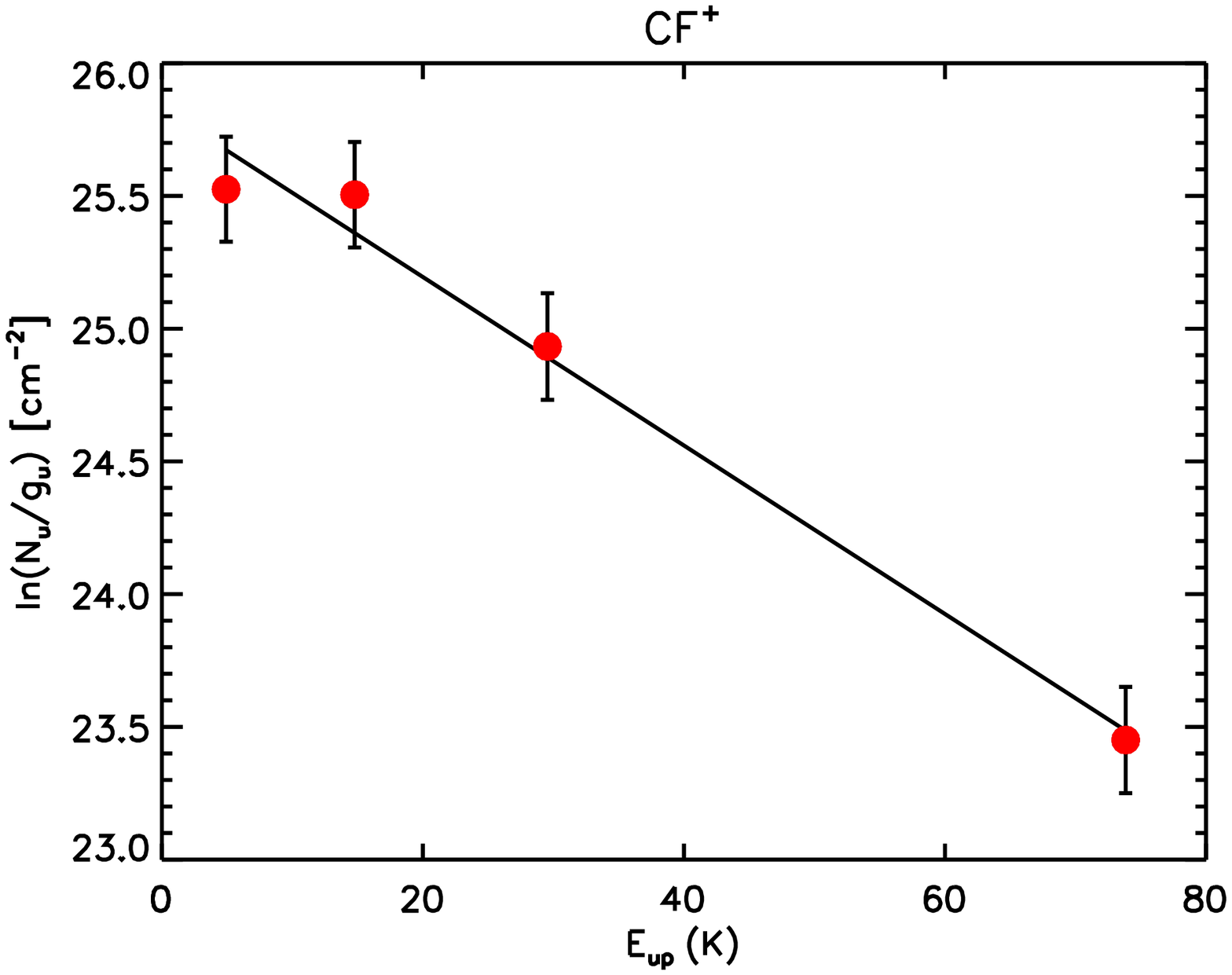} 	    
\caption{Rotation diagram of CF$^+$ including the transitions detected by \citet{neufeld2006}.}
\label{rot_diagram_cfplus}
\end{figure}

We have also detected the CF$^+$ 5-4 transition for the first time, with a line width of $\sim$2 \kms (Fig. \ref{line_profiles_cfplus}). The 3-2, 2-1, and 1-0 transitions of CF$^+$ were previously detected toward the Orion Bar from the ground by \citet{neufeld2006} with beam sizes of HPBW=24$''$, 12$''$, and 21$''$, respectively. The velocity and the width of the $5-4$ transition is consistent with the parameters reported for the other detected transitions by \citet{neufeld2006}. 
One of the positions covered by \citet{neufeld2006}, $\mathrm{05^h35^m22.8^s}$, $-5^\circ25'01''$, is close ($\Delta {\rm{RA}}\sim30''$, $\Delta$Dec$\sim13''$) to our observed position which is within the beam of HIFI at the frequency of the 5-4 transition ($\sim$44.2$''$).
Assuming uniform beam-filling, a single excitation temperature of all four levels and optically thin lines, the measured line intensity is consistent with those detected by \citet{neufeld2006} (Fig. \ref{rot_diagram_cfplus}) and implies a column density of $\sim$2.1$\times$10$^{12}$ cm$^{-2}$ and a rotation temperature of $\sim$32 K. \\
We derived upper limits for other non-detected CF$^+$ and SH$^+$ transitions. We estimated rms noise levels in the averaged spectrum of H- and V-polarizations; 3$\sigma$ upper limits on the integrated line intensities are estimated using (e.g. \citealp{coutens2012})
\[I(3\sigma)~[\mathrm{K~km~s^{-1}}] = 3~{\rm{rms}} \sqrt{2~{\rm{d}}v~FWHM}, \]
where d$v$ is the channel width in \kms and the rms noise level is derived in a velocity range of $\pm$5 \kms around the expected velocity. We use a full width at half maximum (FWHM) of 1.9 \kms for the $^{13}$CF$^+$ 5-4 (488664.3 MHz) and the CF$^+$ 6-5 (615365.6 MHz) upper limits, and FWHM$=$3 \kms for SH$^+$ $N_J=1_1-0_1$ ($\sim$683 GHz). The derived 3$\sigma$ upper limits are listed in Table \ref{table:line_param}. 

\begin{table*}[t]
\begin{savenotes}
\begin{minipage}[!h]{\linewidth}\centering
\caption{The detected CH$^+$, SH$^+$, and CF$^+$ transitions; the Gaussian fit parameters for the velocity-resolved transitions observed with HIFI and PACS; and upper limits for non-detections of other CF$^+$ and SH$^+$ transitions. The CH$^+$ $J=3-2,4-3,5-4$, and $6-5$ transitions are spectrally unresolved from PACS. The parameters from HIFI are based on the average spectrum of H- and V-polarizations, unless otherwise specified.}
\label{table:line_param}
\renewcommand{\footnoterule}{}
\begin{tabular}{llllllllllll}
\hline
\hline
\footnotetext[1]{Based on V-polarization data only.}
\footnotetext[2]{Fixed parameter in the fit.}
\footnotetext[3]{3$\times$rms noise level.}
Line & $\int T_{\rm MB}{\mathrm d}V$& $V_{\rm LSR}$ & $\Delta V$& \hskip+0.2cm $T_{\rm peak}$& rms($T_{\rm{MB}}$)\\
     &               (K km s$^{-1}$)& (km s$^{-1}$) & (km s$^{-1}$) & \hskip+0.2cm (K)& (K)\\
\hline
\hskip+0.2cm CH$^+$ 1$-$0& \hskip+0.2cm 24.9$\pm$0.20& 10.5$\pm$0.02& 5.46$\pm$0.04& \hskip+0.2cm 4.28$\pm$0.15& 0.12\\
\hskip+0.2cm CH$^+$ 2$-$1& \hskip+0.2cm 10.6$\pm$0.20& 10.4$\pm$0.05& 4.57$\pm$0.11& \hskip+0.2cm 2.18$\pm$0.22& 0.23\\
\hskip+0.2cm CH$^+$ 3$-$2& \hskip+0.2cm  2.19$\pm$0.31\\       
\hskip+0.2cm CH$^+$ 4$-$3& \hskip+0.2cm  1.01$\pm$0.19\\       
\hskip+0.2cm CH$^+$ 5$-$4& \hskip+0.2cm  0.42$\pm$0.15\\       
\hskip+0.2cm CH$^+$ 6$-$5& \hskip+0.2cm  0.17$\pm$0.06\\       
\hline
$^{13}$CH$^+$ 1$-$0\footnotemark[1]& \hskip+0.2cm 0.46$\pm$0.11& 11.39$\pm$0.15& 1.30$\pm$0.34& \hskip+0.2cm 0.34$\pm$0.10& 0.15\\             
\hline
\hskip+0.2cm SH$^+$ $N_J=1_2-0_1$ $F=3/2-1/2$& \hskip+0.2cm 0.34$\pm$0.02& 10.92$\pm$0.09& 3.00\footnotemark[2]&  \hskip+0.2cm 0.11$\pm$0.02& 0.02\\
\hskip+0.2cm SH$^+$ $N_J=1_2-0_1$ $F=5/2-3/2$& \hskip+0.2cm 0.57$\pm$0.02& 10.84$\pm$0.06& 3.00\footnotemark[2]& \hskip+0.2cm 0.18$\pm$0.02& 0.02\\
\hskip+0.2cm SH$^+$ $N_J=1_2-0_1$ $F=3/2-3/2$& \hskip+0.2cm 0.14$\pm$0.02& 10.34$\pm$0.24& 3.00\footnotemark[2]& \hskip+0.2cm 0.04$\pm$0.01& 0.02\\ 
\hskip+0.2cm SH$^+$ $N_J=1_1-0_1$ $F=3/2-1/2$& $\leq$0.18& & & $\leq$0.15\footnotemark[3]& \\
\hskip+0.2cm SH$^+$ $N_J=1_1-0_1$ $F=1/2-1/2$& $\leq$0.11& & & $\leq$0.09\footnotemark[3]& \\
\hskip+0.2cm SH$^+$ $N_J=1_1-0_1$ $F=3/2-3/2$& $\leq$0.13& & & $\leq$0.12\footnotemark[3]& \\
\hskip+0.2cm SH$^+$ $N_J=1_1-0_1$ $F=1/2-3/2$& $\leq$0.11& & & $\leq$0.09\footnotemark[3]& \\
\hline
\hskip+0.2cm CF$^+$ 5-4& \hskip+0.2cm 0.20$\pm$0.02& 11.13$\pm$0.11& 1.96$\pm$0.21& \hskip+0.2cm 0.10$\pm$0.01& 0.02\\
\hskip+0.2cm CF$^+$ 6-5& $\leq$0.08& & &   $\leq$0.08\footnotemark[3]& \\
\hline
$^{13}$CF$^+$ 5-4& $\leq$0.05& & & $\leq$0.05\footnotemark[3]& \\
\hline
\end{tabular}
\end{minipage}
\end{savenotes}
\end{table*}
   	
\subsection{Physical conditions traced by CH$^+$ and SH$^+$}
\label{radex}

To estimate molecular column densities, we use the non-LTE radiative transfer code RADEX \citep{vandertak2007}. 
We use H$_2$ as a collision partner for the excitation of CH$^+$, as we expect a significant fraction of hydrogen to be in a molecular form at the observed position. Rates for inelastic collisions between CH$^+$ and H are not available, but are expected to be of the same order of magnitude as the rates for inelastic collisions between CH$^+$ and H$_2$. If most H is in an atomic state, the density used as an input parameter is then the sum of $n$(H) and $n$(H$_2$).
We also include excitation via inelastic collisions between CH$^+$ and electrons, as the importance of excitation by electrons for HF has been recently demonstrated by \citet{vandertak2012}. We apply an electron density of $\sim$10 cm$^{-3}$.
This is justified, if we assume that the electron abundance is determined by the abundance of C$^+$. The column density of C$^+$ is approximately 10$^{18}$ cm$^{-2}$ \citep{ossenkopf2012} and the H$_2$ column density is approximately 10$^{22}$ cm$^{-2}$ (e.g. \citealp{habart2010}, \citealp{vanderwiel2009}). This implies an electron abundance of 10$^{-4}$ and using $n(\mathrm{H}_2)=10^5$ cm$^{-3}$, an electron density of 10 cm$^{-3}$.
In the following we consider both H$_2$ and electron collisions to probe the excitation of CH$^+$ and SH$^+$. 

For CH$^+$ our calculations are based on collision rates from \citet{turpin2010}, for temperatures in the range between 10 K and 200 K, covering transitions up to the 5-4 transition ($E_\mathrm{up}=599.5$ K). These rates have been scaled from CH$^+$~$-$He to CH$^+-$H$_2$ based on \citet{schoier2005}. For electron collisions, we use collision rates from \citet{lim1999} that are available for temperatures between 100 and 15000 K.

Collisions of H$_2$ and electrons are not always inelastic, but may lead to a chemical reaction. In the case of CH$^+$ this is important since the collision rates with H$_2$ and electrons are comparable to the chemical reaction rates for CH$^+$ with H$_2$ and electrons. For example, for CH$^+$-e$^-$ the chemical reaction rate is 9$\times$10$^{-8}$ cm$^3$ s$^{-1}$ for the destruction \citep{woodall2007} and is 6.4$\times$10$^{-7}$ cm$^3$ s$^{-1}$ for the excitation of the $J=1-0$ transition at 1000 K. 
For CH$^+$-H$_2$ the destruction rate is 1.2$\times$10$^{-9}$ cm$^3$ s$^{-1}$ \citep{woodall2007} and an excitation rate is 1.1$\times$10$^{-10}$ cm$^3$ s$^{-1}$ for the $J=1-0$ transition at 100 K.
Therefore, we consider the chemical formation and destruction rates in the statistical equilibrium calculation (e.g. \citealp{vandertak2007}). The statistical equilibrium for states $i=1$ - $N$ of energy $E_i$ that is solved using the RADEX code is given by the time-independent rate equations
\begin{equation}
\nonumber
{{dn_i}\over{dt}} = \sum_{j\neq i}^N n_j P_{ji} - n_i \sum_{j\neq i}^N P_{ij} = {\cal F}_i - n_i {\cal D}_i \;\;\;{\rm cm}^{-3}\;{\rm s}^{-1},
\end{equation}
where 
\begin{eqnarray} 
P_{ij} & = & A_{ij} + B_{ij} \bar{J} + C_{ij} \;\;\;(E_i>E_j) \nonumber \\
& = & B_{ij}\bar{J} + C_{ij} \;\;\;(E_i<E_j) \nonumber
\end{eqnarray}
and $A_{ij}$ and $B_{ij}$ are the Einstein coefficients, $\bar{J}$ is the mean intensity at the frequency of transition $i\to j$, $C_{ij}$ is the sum over all collision partners of the rates of inelastic, collision-induced transitions $i\to j$, $n_i$ is the number density (cm$^{-3}$) of molecules in level $i$, and ${\cal D}_i$ is the rate of destruction of the molecule in level $i$. When detailed knowledge of the state-specific formation process is lacking, the formation rate into level $i$ is expressed as a Boltzmann distribution over all states at an effective formation temperature $T_f$
\begin{equation}
\nonumber
{\cal F}_i \propto g_i \exp(-E_i/kT_f),
\end{equation}
where $g_i$ is the statistical weight of level $i$. When the destruction rate can be estimated, as for CH$^+$ here, then the total formation rate is normalized so that the total number density of molecules is consistent with its column density and the density of hydrogen in steady state.

We assume that there is a balance between the formation and destruction of CH$^+$. 
To simulate the chemical pumping effect described above, i.e. the effect of destruction and subsequent formation of CH$^+$ in excited levels, we add an artificial level to the CH$^+$ level system, representing the dissociated state, that is populated with a rate equivalent to the reaction rate of CH$^+$-H$_2$ and CH$^+$-e$^-$ \citep{woodall2007}. 
On the formation of CH$^+$ through the reaction of C$^+$ with vibrationally excited H$_2$, the re-population from the dissociated level follows a Boltzmann distribution with a formation temperature of $T_f=9920~{\rm{K}} - 4560~{\rm{K}}=5360~{\rm{K}}$, where 4560 K is the required energy input for the endothermic CH$^+$ production and 9920 K is the average energy of the vibrationally excited H$_2$ levels following the 2-level approximation introduced by \citet{rollig2006} where the full 15-level system was replaced by the energetically equivalent 2-level system that provides the same total vibrational heating.

\begin{figure}[!h]
\centering
\includegraphics[width=9 cm, trim=0cm 0cm -0.5cm 0,clip=true]{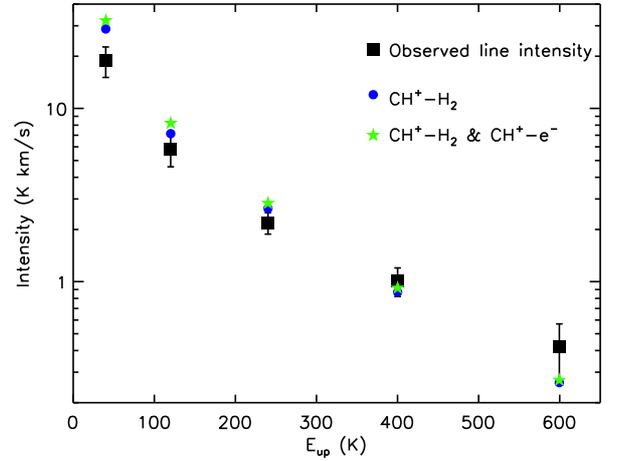}     
\caption{Output of the RADEX models corresponding to a model with $N$(CH$^+$)$=9\times10^{14}$ cm$^{-2}$, $n$(H$_2$)$=10^5$ cm$^{-3}$, $n$(e$^-$)$=10$ cm$^{-3}$, $T_{\rm{kin}}=500$ K. The blue symbols correspond to a model with excitation via collisions with H$_2$, the green symbols to a model with excitation via H$_2$ and electron collisions.}
\label{chp_int_radex}
\end{figure}

Figure \ref{chp_int_radex} shows the intensity predictions of two RADEX models with parameters in the range that can be expected for the Orion Bar. The error bars correspond to a 10\% calibration error and a 10\% error from obtaining integrated intensities by Gaussian fitting for the transitions observed with HIFI. 
The error bars corresponding to the PACS data are dominated by uncertainty in the PSF correction and are estimated as explained in Sect. \ref{obs}.
Both electron collisions and H$_2$ collisions suggest a kinetic temperature well above the average value (85 K) inferred for the interclump medium in the Orion Bar. 
Taking the formation pumping and collisional excitation described above into account, we find reasonable fits to the observed line-intensity distribution up to the 5-4 transition (the energy range for which the collision rates are available) with $N$(CH$^+$)$=9\times10^{14}$ cm$^{-2}$, $T_{\rm{kin}}=500$ K, and $n$(H$_2$)=10$^5$ cm$^{-3}$. The intensity of the 5-4 transition can be better reproduced with a kinetic temperature of $T_{\rm{kin}}=1000$ K. 
Temperatures between 500 K and 1000 K are expected near the edge of the cloud where the observations used in this paper have been taken. 
Assuming an electron density of $n_{\rm{e}}=10$ cm$^{-3}$, electron collisions mostly affect the two lowest$-J$ transitions for $T_\mathrm{kin}=500$ K with an 10-13\% increase in the intensities. 

\begin{figure}[!h]
\centering
\includegraphics[width=9 cm, trim=0cm 0cm -0.5cm 0,clip=true]{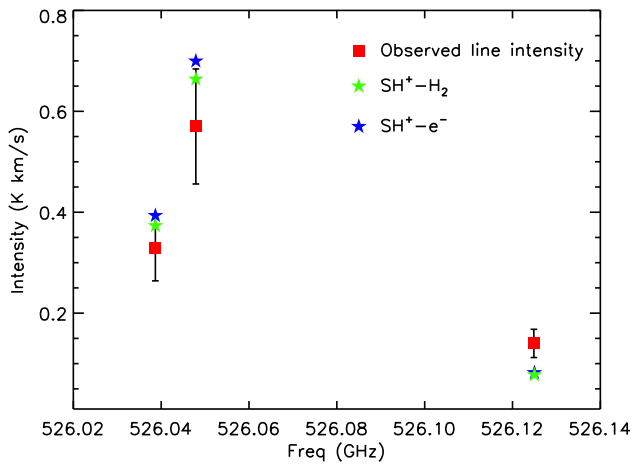} 	    
\caption{Output of the RADEX models for H$_2$ and electron collisions compared to the observed line intensities of SH$^+$ for a model with $N(\mathrm{SH^+})=10^{13}$ cm$^{-2}$, $n(\mathrm{H_2})=10^6$ cm$^{-3}$, $n({\rm{e}}^-)=10$ cm$^{-3}$, and $T_{\rm{kin}}=200$ K.}
\label{shp_int_radex}
\end{figure}

In the case of SH$^+$, calculations for collisions with H$_2$ do not exist, so we use scaled radiative rates (J. Black, private communication) for a temperature range of 10-5000 K. We also use collision rates for electron-impact collisions calculated in the Coulomb-Born approximation (J. Black, private communication), for a temperature range of 10-1000 K. Figure \ref{shp_int_radex} shows the best fit models over-plotted on the observed line intensities. The error bars correspond to 20\% of the observed line intensities, including calibration error and the error introduced by the estimation of the integrated intensities using a Gaussian fit. 
We consider lower kinetic temperatures and higher volume densities than in the case of CH$^+$, given that the line width suggests an origin from denser material. At the position of the CO$^+$ peak, warm ($T_\mathrm{kin}\sim160-220$ K) and dense (10$^{6-7}$ cm$^{-3}$) condensations have been suggested to explain the OH emission.
Using SH$^+$-H$_2$ collisions, a model with 10$^6$ cm$^{-3}$, $T_{\rm{kin}}=200$ K, and $N$(SH$^+$)=10$^{13}$ cm$^{-2}$ gives a reasonable fit to the observed line intensities (Fig. \ref{shp_int_radex}). On Fig. \ref{shp_int_radex} we show a model with the same parameters, which includes electron collisions, assuming an electron density of $n_{\rm{e}}=10$ cm$^{-3}$. 
In both of these models the excitation temperatures are low (8.3-10.4 K) and the lines are optically thin ($\tau \sim0.02-0.2$). 

\section{The formation of CH$^+$ and SH$^+$ via H$_2$ vibrational excitation}

In this section we investigate the role of H$_2$ vibrational excitation for the formation of CH$^+$ and SH$^+$. We discuss alternative explanations in Sect. \ref{discussion}.

\subsection{Estimate based on an analytic approximation} 
 
Testing if H$_2$ vibrational excitation can drive CH$^+$ and SH$^+$ formation in an environment with a given radiation field and physical parameters requires a detailed modeling with a PDR code, with information on the chemical network and physical processes that affect the level populations of vibrationally excited H$_2$.
A first indication can also be given using a simple analytic method to describe H$_2$ vibrational heating with a two level approximation \citep{rollig2006}. The H$_2$ vibrational heating rate can be computed among all 15 vibrational levels in the ground electronic state, but neglecting the rotational structure, based on Equation C.2 in \citet{rollig2006}. A two-level system can be defined that results in the same vibrational heating rate as the full, 15-level system (Equation C.3, \citealp{rollig2006}).
The vibrationally excited `virtual' level has an upper level energy of $\Delta E_{\rm{eff}}$=9920 K. 
The total rate for populating this vibrationally excited level is 
$k_{0,1} = P_1~\chi$,
where $\chi$ is the radiation field in Draine units \citep{draine1978} and $P_1=6.9\times10^{-10}~\rm s^{-1}$ is the formation rate of vibrationally excited H$_2$ for the defined level of a radiation field of $\chi=1$.
The de-excitation of this vibrationally excited level is via spontaneous emission, dissociation by the UV radiation field and by collisional de-excitation.
The coefficient for spontaneous decay is $A_{\rm{eff}}=1.9\times10^{-6}~\rm s^{-1}$. The collisional de-excitation scales as $n_{\rm{gas}} \gamma_{\rm{eff}}$ with a rate coefficient of $\gamma_\mathrm{eff}=5.4\times10^{-13}\sqrt{T}~{\rm{s}}^{-1}~{\rm{cm}}^{-3}$. The dissociation rate is $\chi \times D_{\rm{eff}}$, where $D_{\rm{eff}}=4.7 ~10^{-10} \rm s^{-1}$. 
These effective coefficients $A_{\rm{eff}}$, $\gamma_\mathrm{eff}$, and $D_{\rm{eff}}$ for the defined 2-level system as well as the energy of the defined vibrationally excited level $\Delta E_{\rm{eff}}$ are obtained by considering different asymptotic values of the density $n$ and the radiation field $\chi$.
By neglecting dissociation, the population of the vibrationally excited level is dependent on the formation rate of vibrationally excited H$_2$ as well as on the spontaneous decay and collisional de-excitation rates. Calculating with $n_{\rm{gas}}=10^5~{\rm{cm}}^{-3}$ and $T=500$ K for the collisional de-excitation rate, there is a balance between these processes for a radiation field of $\chi \sim 5\times10^{3}$, which gives an expected lower limit on the radiation field, above which H$_2$ vibrational excitation is expected to be efficient enough to drive the formation of CH$^+$ and SH$^+$.
Even though neglecting dissociation introduces an additional $\sim$10\% error, this calculation shows that for the radiation field in the Orion Bar ($1-4\times10^4$ in Draine units), there is a large percentage of vibrationally excited H$_2$ to react with C$^+$ and form CH$^+$. This has been observed by \citet{vanderwerf1996} and \citet{walmsley2000} and has already been noted for the formation of OH through the O + H$_2$ $\rightarrow$ OH + H reaction by \citet{goicoechea2011}.
In the following section we test this idea with a more accurate approach, using PDR models.

\subsection{CH$^+$ formation}

We use the 1.4.4 version of the Meudon PDR code (\citealp{lepetit2006}, \citealp{goicoechea2007}, \citealp{lebourlot2012}) to model the observed CH$^+$ line intensities. This version includes the Langmuir Hinshelwood and Eley-Rideal mechanisms to describe the formation of H$_2$ on grain surfaces. 
The chemical pumping effect of destruction and formation on CH$^+$ level populations is taken into account in addition to collisional excitation and de-excitation in the Meudon code (e.g. \citealp{gonzalezgarcia2008}).
The Meudon code treats CH$^+$ formation as described in \citet{agundez2010}
\begin{equation}
\label{chp_form_1}
\mathrm{H_2(j = 0-7) + C^+ \xrightarrow{k_1} CH^+ + H}
\end{equation}
\begin{equation}
\mathrm{H_2 (v=1) + C^+ \xrightarrow{k_2} CH^+ + H}
\end{equation}
In reaction \ref{chp_form_1}, H$_2$ rotational levels up to $J$=7 are used, which has an energy, $E_7$=4586.4 K which is close to the activation barrier of the $\mathrm{H_2 + C^+ \rightarrow CH^+ + H}$ reaction. We take into account the $v=1$ vibrational level only because its energy ($\sim$5987 K) is enough to overcome the activation barrier of the CH$^+$ formation reaction. The formation rates are $k_1=1.58 \times 10^{-10} \exp(-[4827 - E_j/k]/T )$ based on \citet{gerlich1987} and $k_2=1.6 \times 10^{-9}$ \citep{hierl1997}.

We use isobaric models for typical conditions for the Orion Bar with pressures in the range between 5$\times$10$^7$ cm$^{-3}$ K and 2$\times$10$^8$ cm$^{-3}$ K, corresponding to $T_{\rm{kin}}\sim500$ K (RADEX models) and n$\sim$10$^5$ cm$^{-3}$; and $T_{\rm{kin}}\sim1000$ K (RADEX models) and n$\sim$2$\times$10$^5$ cm$^{-3}$ (typical interclump medium density, e.g. \citealp{simon1997}); respectively. 
We apply a radiation field on the side where the cloud is illuminated from in the range between $\chi_{\rm{front}}$=10$^4$ and 3$\times$10$^4$ in Draine units \citep{draine1978}. 
We run the models up to a depth equivalent to a visual extinction of $A_{\rm{V}}\sim10$ mag. At the back side of the cloud (at $A_{\rm{V}}\sim10$ mag), we use a radiation field 1000 times below that on the front $\chi_{\rm{back}}=\chi_{\rm{front}}/1000$. We adopt a cosmic-ray primary ionization rate of $\zeta=2\times10^{-16}$ s$^{-1}$ per H$_2$ molecule suitable for the dense ISM \citep{hollenbach2012}.

\begin{figure}[!h]
\centering
\includegraphics[width=8.5 cm,trim=1.0cm 0cm 0cm 0cm,clip=true]{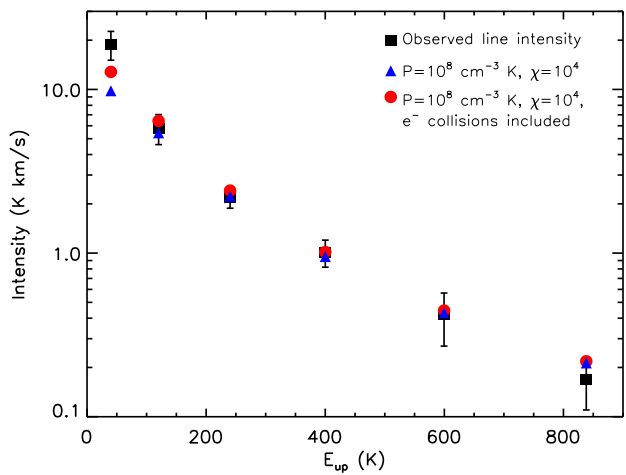}
\caption{CH$^+$ line intensities as a function of $E_{\rm{u}}$, comparison between the observed line intensities and the predictions of an isobaric PDR model with a pressure of $10^8$ cm$^{-3}$ K for a radiation field of $\chi=10^4$.}
\label{chp_int_meudon}
\end{figure}

Figure \ref{chp_int_meudon} shows the results of a model for a pressure of $10^8$ cm$^{-3}$ K and a radiation field of $\chi_{\rm{front}}$=1$\times$10$^4$, consistent with the radiation field near the ionization front of the Orion Bar.
An inclination of 60$^\circ$ was used to extract the line intensities, because of uncertainties in the computation of the line intensities in a 1D model above this value \citep{gonzalezgarcia2008}. This is reasonably close to the model with 75$^\circ$ inclination suggested to explain the geometry of the Bar (e.g. \citealp{melnick2012}).
The model with an inclination of 60$^\circ$ reproduces the observed CH$^+$ line intensities within a factor of 2 for the J=1-0 transition, and with an accuracy of 20\% for the other transitions.
Our RADEX calculations show the possible importance of electron collisions in the excitation of CH$^+$. Therefore, to probe the effect of electron collisions on the excitation of CH$^+$, we implemented CH$^+$-e$^-$ collisions in the Meudon code. The models for P=$10^8$ cm$^{-3}$ K and $\chi=10^4$ are shown in Fig. \ref{chp_int_meudon}. Including electrons in the excitation of CH$^+$, the model reproduces the observed line intensities with an accuracy of $\sim30$\%. Including electron collisions affects mostly the two lowest$-J$ transitions. The predicted intensity of the $J=1-0$ transition increases by $\sim$22\%, and the intensity of the $J=2-1$ transition increases by $\sim$18\% after including electron collisions.

\begin{figure}[!h]
\centering
\includegraphics[width=6.7 cm, angle=-90,trim=0cm 0cm 0 0,clip=true]{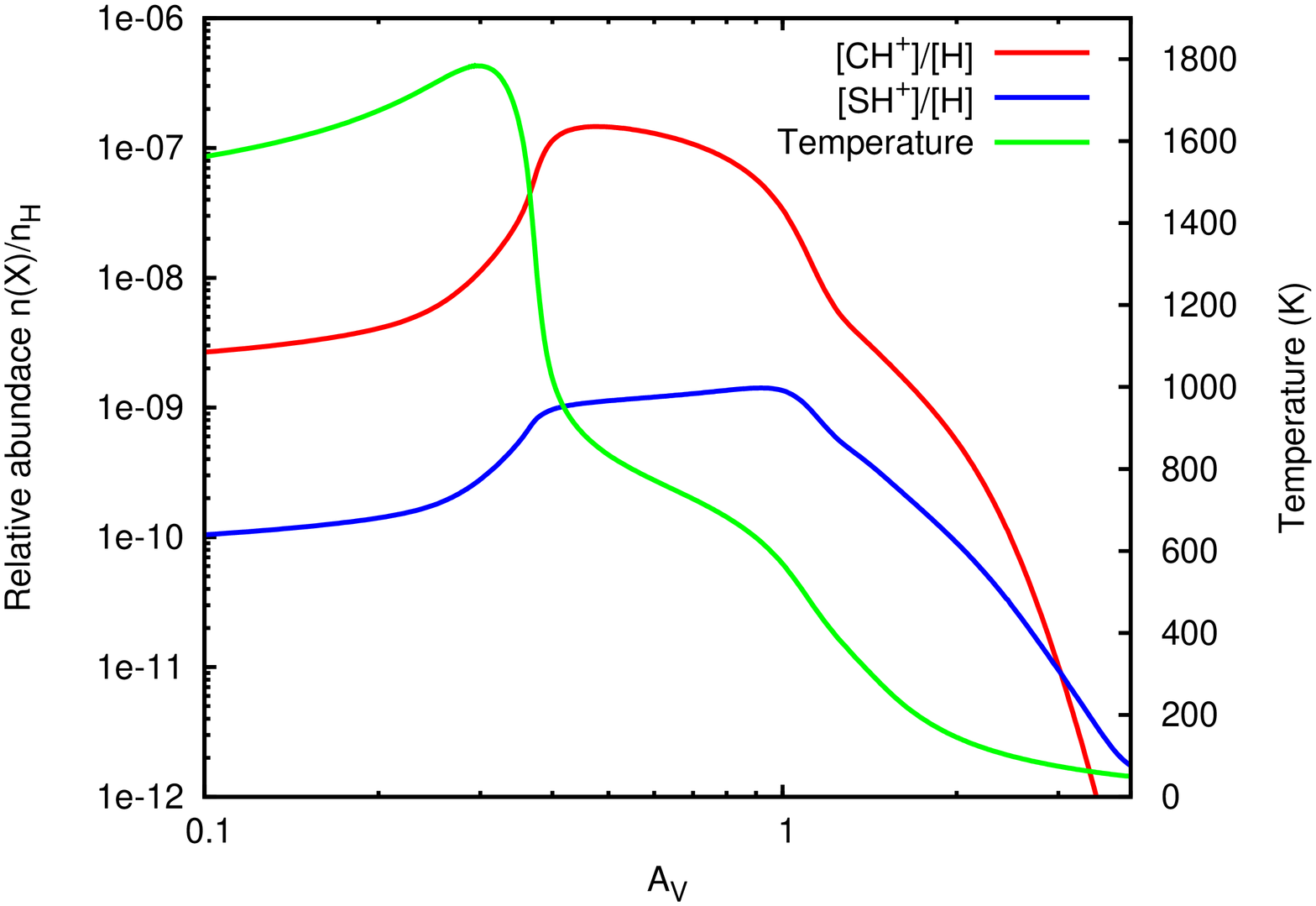} \\
\hskip-0.9cm
\includegraphics[width=5.7 cm, angle=-90,trim=0cm 0cm 0cm 1cm,clip=true]{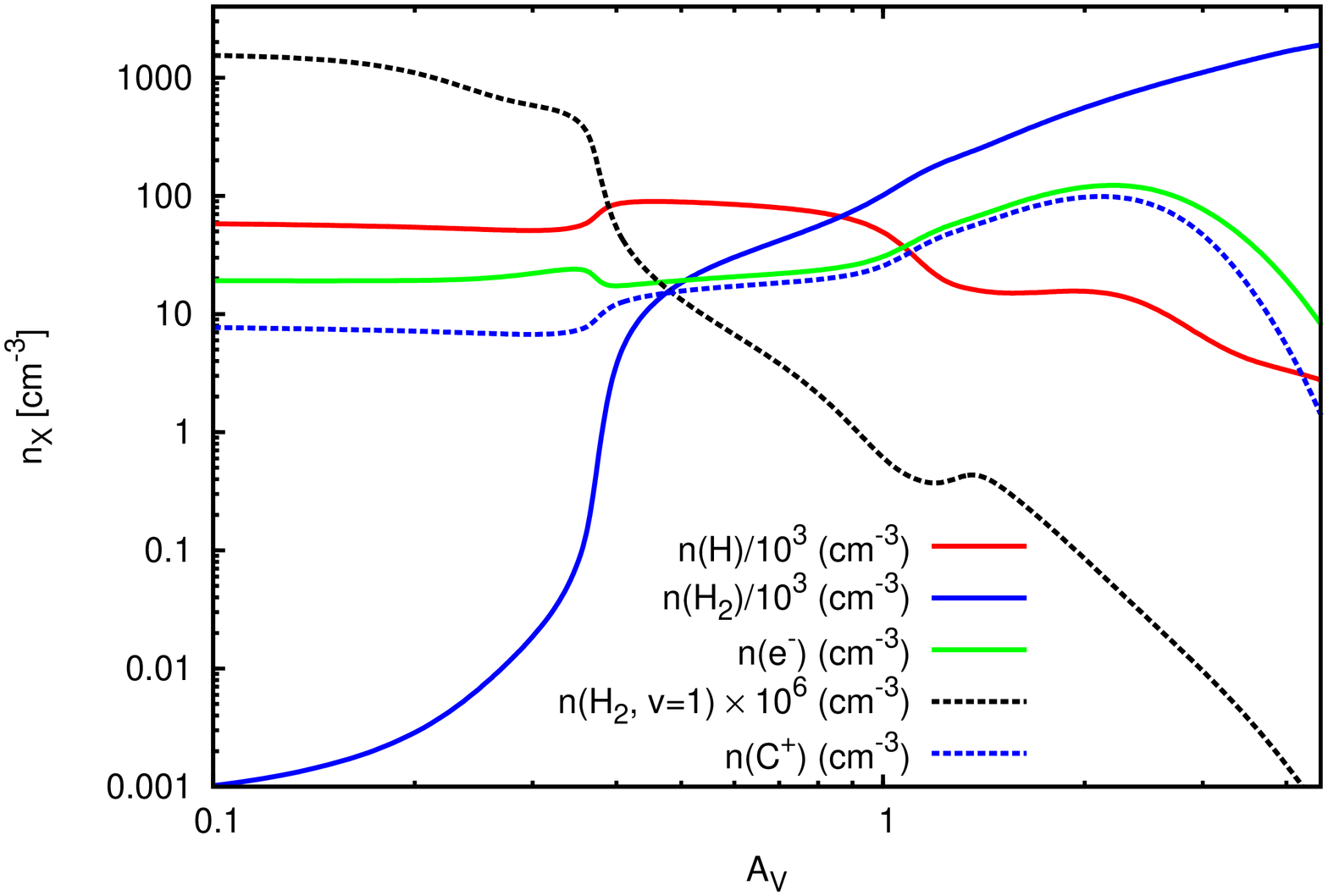}
\caption{\emph{Top panel}: CH$^+$ and SH$^+$ abundances and the gas kinetic temperature as a function of $A_{\rm{V}}$ for a pressure of $P=10^8$ cm$^{-3}$ K and $\chi=10^4$. \emph{Bottom panel:} H, H$_2$, H$_2$ (v=1), e$^-$, and C$^+$ densities.}
\label{abundance}
\end{figure}

The CH$^+$ abundance profile corresponding to this model is shown in Fig. \ref{abundance} together with the gas temperature in the region where CH$^+$ abundances peak.
The model predicts that CH$^+$ forms near the surface of the PDR ($A_{\rm{V}}<1$) at high temperatures ($T\sim500-1000$ K), consistent with the predictions by \citet{agundez2010} and with our RADEX calculations.
Though the best fitting models predict abundances to peak near the surface of the cloud at low $A_{\rm{V}}$, the PACS observations of excited CH$^+$ used in this paper show a spatial extension along the area covered by PACS ($47''\times47''$), as shown in Fig. \ref{pacs_cso_maps_CHp32}. SPIRE observations of the $J=1-0$ transition (\citealp{naylor2010}, \citealp{habart2010}) show extended CH$^+$ emission over a $\sim200''\times200''$ region centered on the $\alpha_\mathrm{J2000}=\rm{05^h35^m22.83^s}$, $\delta_\mathrm{J2000}=-05^\circ 24'57.67''$ position.
The CH$^+$ $J=1-0$ emission mapped with HIFI was found to extend over a large region covering the OMC-1 cloud (Goicoechea et al., in preparation).
One possibility is that the known clumpiness of the Orion Bar extends over a large volume and creates multiple PDR surfaces. Alternative explanation is that the extended CH$^+$ emission seen toward the region is the result of a not completely edge-on PDR that is tilted to the line of sight.
Models with lower pressures under-predict the observed line intensities. 
For example, a model with a pressure of 5$\times$10$^7$ K cm$^{-3}$ underpredicts the line intensities with a factor of $\sim$4.
     
\subsection{SH$^+$ formation}
\label{shp_form}

SH$^+$ forms via a similar reaction to CH$^+$, however, with an endothermicity about twice as high, $\Delta$E=9860 K. Since state-to-state formation rates are not available for the $\mathrm{H_2 + S^+ \rightarrow SH^+ + H}$ reaction, we use the rates of the $\mathrm{H_2 + C^+ \rightarrow CH^+ + H}$ reaction as an approximation, taking into account H$_2$ rotational levels up to $E_{11}=10261.8$ K and H$_2$ in the v=1 state up to $E=10341.5$ K.
This is a reasonable assumption, since the total rates of the reactions for CH$^+$ and SH$^+$ formation via C$^+$ + H$_2$ and S$^+$ + H$_2$ are of the same order of magnitude \citep{woodall2007}.
To account for the higher activation barrier, we use $k_{1,\mathrm{mod}}=1.58 \times 10^{-10} \exp(-[9860 - E_j/k]/T )$. 
We use the 1.4.4 version of the Meudon code (\citealp{lepetit2006}, \citealp{goicoechea2007}, \citealp{lebourlot2012}), where we introduce the SH$^+$ formation described above and use scaled radiative rates (J. Black, private communication) and electron-impact collisions calculated in the Coulomb-Born approximation (J. Black, private communication) for the excitation of SH$^+$.

With these assumptions, our best fit Meudon PDR model for CH$^+$ ($P=10^8$ cm$^{-3}$ K, $\chi=10^4$) underpredicts the absolute intensities of the observed SH$^+$ transitions by a factor of $\sim$3.5 for the $F=5/2\rightarrow3/2$ and $F=3/2\rightarrow1/2$ transitions, and by a factor of 6.5 for the $F=3/2\rightarrow3/2$ transition. Owing to the uncertainty in the formation rates, this agreement may be reasonable and suggests that like CH$^+$, SH$^+$ can also be formed via H$_2$ vibrational excitation in warm and dense PDRs. 
It may also suggest that SH$^+$ originates in a higher-pressure medium compared to CH$^+$, which would explain the difference in the observed linewidths. The SH$^+$ abundances corresponding to this model are shown in Fig. \ref{abundance}. SH$^+$ abundances, like CH$^+$ abundances, peak near the surface of the cloud at $A_{\rm{V}}\lesssim1$ at high temperatures (500-1000 K). The SH$^+$/CH$^+$ abundance ratio in this region is between 0.01 and 0.1. This, however, is a lower limit on the SH$^+$/CH$^+$ abundance ratio, since our model underestimates the SH$^+$ line intensities.

\begin{figure}[!h]
\centering
\hskip-0.3cm
\includegraphics[width=7.5 cm,trim=0.5cm 0cm 0cm 0cm,clip=true]{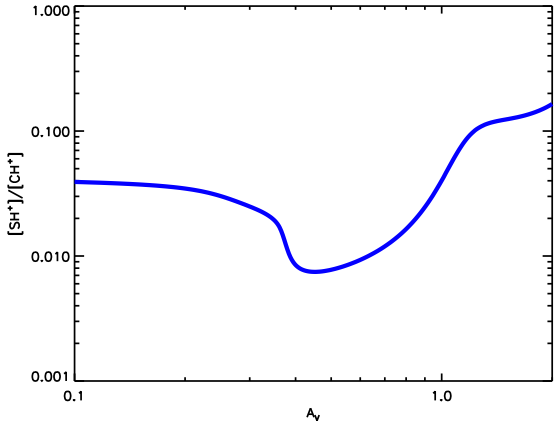}
\caption{Abundance ratios of CH$^+$ and SH$^+$ in our best fit model ($\chi=10^4$, $P=10^8$ cm$^{-3}$ K) in the warm surface region ($A_{\rm{V}}<2$) as a function of depth into the cloud.}
\label{chp_shp_ab_ratio}
\end{figure}

\section{The destruction of CH$^+$ and SH$^+$}
\label{destruction}

Figure \ref{chp_shp_ab_ratio} shows the abundance ratio of CH$^+$ and SH$^+$ predicted by our best-fit model in the region where CH$^+$ and SH$^+$ abundances peak. To understand these abundance ratios, it is essential to study the destruction of CH$^+$ and SH$^+$.
CH$^+$ destruction can follow four main paths in the probed temperature and density regime
\[\mathrm{H + CH^+ \xrightarrow{k_{H,CH^+}} H_2 + C^+}\]
\[\mathrm{H_2 + CH^+ \xrightarrow{k_{H_2,CH^+}} H + CH_2^+}\]
\[\mathrm{e^- + CH^+ \xrightarrow{k_{e^-,CH^+}} H + C}\]
\[h\nu + \mathrm{CH^+ \xrightarrow{k_{h\nu, CH^+}} C + H^+}\]
Unlike CH$^+$, SH$^+$ does not react with H$_2$ at the given physical conditions, as the reaction rate is orders of magnitude lower than that of reactions with electrons and H. Therefore the most important destruction paths are photodissociation as well as chemical reactions with H and electrons
\[\mathrm{H + SH^+ \xrightarrow{k_{H,SH^+}} H_2 + S^+}\]
\[\mathrm{e^- + SH^+ \xrightarrow{k_{e^-,SH^+}} H + S}\]  
\[h\nu + \mathrm{SH^+ \xrightarrow{k_{h\nu, SH^+}} H + S^+}\]
The chemical reaction rates for these reactions are based on \citet{woodall2007} and are summarized in Table \ref{table:rates}. 

\begin{table}[ht]
\begin{savenotes}
\begin{minipage}[!h]{\linewidth}\centering
\caption{Rates corresponding to the main destruction paths of CH$^+$ and SH$^+$, based on \citet{woodall2007}.}
\label{table:rates}
\renewcommand{\footnoterule}{}
\begin{tabular}{lll}
\hline
Reaction& Rate& Temperature regime\\
& (cm$^{3}$ s$^{-1}$)& (K)\\
\hline
$k_{\rm{H,CH^+}}$& $7.5\times10^{-10}$& $10-41000$\\ 

$k_{\rm{H_2,CH^+}}$& $1.2\times10^{-9}$& $10-41000$\\ 

$k_\mathrm{e^-,CH^+}$& $1.5\times10^{-7}\times(T/300)^{-0.42}$& $10-300$\\

$k_{\rm{H,SH^+}}$& $1.10\times10^{-10}$& $10-41000$\\ 

$k_{\rm{e^-,SH^+}}$& $2.0\times10^{-7}\times(T/300)^{-0.5}$& $10-300$\\

\hline

Reaction& Rate& Temperature regime\\
& (s$^{-1}$)& (K)\\

\hline

$k_\mathrm{h\nu,CH^+}$& $2.50\times10^{-10} \exp(-2.5 A_{\rm{V}})$& $10-41000$\\ 

$k_\mathrm{h\nu,SH^+}$& $3.00\times10^{-10} \exp(-1.8 A_{\rm{V}})$& $10-41000$\\

\hline
\end{tabular}
\end{minipage}
\end{savenotes}
\end{table}

At a depth equivalent to a visual extinction of A$_{\rm{V}}\sim$0.5, which is in the region where the CH$^+$ and SH$^+$ abundances peak, and at a temperature of $\sim$830 K, most CH$^+$ (73.2\%) is destroyed via collisions with H, while 24.3\% is destroyed via collisions with H$_2$ and 2.1\% is destroyed via collisions with electrons. Photodissociation is negligible in this regime, since it is responsible for 0.4\% of the destructions of CH$^+$.
Deeper in the cloud, at a depth equivalent to $A_{\rm{V}}\sim1$, and at a gas temperature of $\sim$570 K, most CH$^+$ (74.9\%) is destroyed by collisions with H$_2$, while small percentages of CH$^+$ are destroyed via collisions with H (22.8\%), with electrons (2.2\%), and by photodissociation (0.6\%). 

At a depth of $A_{\rm{V}}\sim0.5$, most SH$^+$ ($\sim$77.2\%) is destroyed via reactions with H and a smaller percentage is destroyed via reactions with electrons ($\sim$18.2\%) and by photodissociation ($\sim$4.4\%).
At a depth corresponding to $A_{\rm{V}}\sim1$, SH$^+$ is almost equally destroyed by reactions with H ($\sim$53.3\%) and electrons ($\sim$43.7\%). A smaller percentage of SH$^+$ is destroyed via photodissociation ($\sim$2.4\%).

At $A_{\rm{V}}\sim1$ and deeper, the SH$^+$ and CH$^+$ abundance ratio becomes higher than 0.1. 
Deeper in the cloud than $A_\mathrm{V}\sim1$ CH$^+$ abundances decrease more rapidly than SH$^+$ abundances.
The [SH$^+$]/[CH$^+$] abundance ratio varies in the range between 0.01 (at $A_{\rm{V}}\sim0.2$) and $>$0.1 (at $A_{\rm{V}}\sim1$) and indicate that the destruction of CH$^+$ becomes more efficient as a function of the depth into the cloud where H$_2$ takes over as the most important destruction partner, while with the decrease of atomic hydrogen density SH$^+$ becomes more abundant because electrons are a less efficient destruction partner. While in diffuse clouds the abundance ratios of [SH$^+$]/[CH$^+$] trace the importance of shocks (e.g. \citealp{menten2011}) and properties of turbulent dissipation regions (e.g. \citealp{godard2012}), in clouds exposed to high UV irradiation these abundances are sensitive to the abundance ratios of H, H$_2$, and electrons as a function of depth into the cloud, which are determined by the radiation field strength of the irradiation source. 

\section{Discussion}
\label{discussion}

We have analyzed six rotational transitions of CH$^+$ and three transitions of SH$^+$ and reported the first detection of the $J=5-4$ transition of CF$^+$. 
We have shown that electron collisions affect the excitation of SH$^+$ and CH$^+$, especially the lowest$-J$ transitions. We have also shown the importance of taking reactive collisions into account in the case of CH$^+$ excitation. 
We have confirmed, both by an analytic approximation and by more detailed PDR modeling, that CH$^+$ formation is driven by H$_2$ vibrational excitation, unlike in the case of diffuse environments with lower UV radiation fields. SH$^+$ is also likely to form via H$_2$ vibrational excitation, although the lack of information on the exact state-to-state formation rates introduces an extra uncertainty in the models.  

\subsection{The formation of CH$^+$ and SH$^+$}

Spatially extended vibrationally excited H$_2$ emission was detected in the Orion Bar before the launch of Herschel (\citet{vanderwerf1996}, \citet{walmsley2000}), already indicating an importance in the chemistry of species that react with H$_2$. Using Herschel, CH$^+$ was detected in the Orion Bar by \citet{naylor2010} and \citet{habart2010}, based on SPIRE maps of the 1-0 transition. These observations show extended CH$^+$ 1-0 emission in the Orion Bar as well as in the OMC-1 cloud (\citealp{naylor2010}, Morris, P.; priv. comm., Goicoechea, J.; priv. comm.).
\citet{naylor2010} argue that the large spatial extent of CH$^+$ into regions of low $A_{\rm{V}}$ suggests the importance of the formation via H$_2$ vibrational excitation. 
Our observations extend these studies, since the additional observed transitions up to $J$=6-5 provide additional evidence on the importance of the formation via vibrationally excited H$_2$.
An origin of CH$^+$ in the warm surface regions of the PDR is also confirmed by \citet{goicoechea2011} who found a spatial correlation between excited OH $^2\Pi_{3/2}$ $J=7/2^- \rightarrow 5/2^+$ ($\sim$84.6 $\mu$m, observed with PACS) and CH$^+$ 3-2 emission, and that OH originates in the surface region ($A_{\rm{V}}<1$) of a high pressure gas component ($10^8-10^9$ K cm$^{-3}$).
The formation and excitation of CH$^+$ in the Orion Bar is similar to that in the envelope of the high-mass protostar AFGL 2591, as its CH$^+$ emission can be explained to originate in the FUV-irradiated outflow-walls (\citealp{bruderer2010}). Another region where CH$^+$ formation is driven by the strong FUV radiation field and can be explained by H$_2$ vibrational excitation is the protoplanetary disc HD 100546 \citep{thi2011}, where CH$^+$ emission mostly originates in the outer disc and the disc surface in warm gas ($T_{\rm{gas}}>400$ K).

Other explanations for the formation of CH$^+$ applicable to the diffuse ISM include shocks (e.g. \citealp{pineaudesforets1986}). \citet{tielens1993} have investigated that shocks do not contribute to the chemistry of the Orion Bar; therefore, we consider this scenario unlikely.  
Another scenario for CH$^+$ formation that has been successful in reproducing CH$^+$ abundances for the diffuse interstellar medium is the dissipation of turbulence \citep{godard2009,godard2012}. Though the CH$^+$ 1-0 and 2-1 transitions have broader line widths than most dense gas tracers in the Orion Bar, most of the lines detected in the Orion Bar are narrow (2-3 \kms); therefore, we find it unlikely that turbulence plays a role in the chemistry of species detected in the Orion Bar. 

Unlike CH$^+$, SH$^+$ has not been observed in a large variety of regions since its recent discovery in absorption toward Sagittarius B2 \citep{menten2011}. A recent study by \citet{godard2012} probes CH$^+$ and SH$^+$ in absorption in the diffuse interstellar medium toward high-mass star-forming regions, suggesting a common origin for the formation and excitation of these ions, based on their observed linewidth-distributions and on comparison with MHD shock models. However, in the diffuse ISM, SH$^+$ and CH$^+$ abundances are influenced by the dissipation of turbulence. SH$^+$ has also been detected in emission in the high-mass star-forming region W3 IRS5 (\citealp{benz2010}), which represents a region with physical conditions comparable to the Orion Bar, where the UV-radiation of the embedded protostars drives the chemistry of SH$^+$. 

\subsection{CH$^+$ and SH$^+$ as tracers of the warm PDR surface}

Though CH$^+$ and SH$^+$ most likely form via the same process and originate in the warm surface region of the PDR, a significant difference between CH$^+$ and SH$^+$ emission is suggested by the difference in the observed line widths. 
While the observed line width of SH$^+$ ($\Delta v \sim$ 3 \kms) is closer to that of dense gas tracers ($\Delta v \sim$ 2-3 \kms), the width of the CH$^+$ $J=1-0$ and 2$-$1 transitions ($\Delta v \sim$ 5 \kms) is similar to that of HF ($\Delta v \sim$ 4.9 \kms, \citealp{vandertak2012}) and C$^+$ ($\Delta v \sim$ 3.8 \kms), tracers of the interclump medium.
The C91$\alpha$ carbon recombination line was observed with the VLA with a width of 2-2.5 km s$^{-1}$ \citep{wyrowski1997}. It was found to match the H$_2$ [1-0 S(1)] distribution \citep{vanderwerf1996} and its radial velocity was found to be consistent with that of H$_2$ pure rotational lines H$_2$ v=0-0 S(1), S(2), and S(4) \citep{allers2005}. 
The $^{13}$C$^+$ lines have a slightly larger width of 2.5-2.8 km s$^{-1}$, compared to that of the C91$\alpha$ line. The larger width of the [C{\sc{ii}}] 158 $\mu$m line compared to the $^{13}$C$^+$ lines can be a result of optical depth broadening of the C$^+$ line (with an optical depth of 2-3). 
However, the C$^+$ line is also broader near the edge of the Bar, where the column density of material is lower, as are the line optical depths.
In addition, the recombination line intensity is sensitive to the square of the electron density, while the fine structure line is sensitive to the local density only. Therefore the difference in line profiles of the C91$\alpha$ and [C{\sc{ii}}] 158 $\mu$m lines outside opacity broadening may also be related to gradients in the beam and along the line of sight, with the denser material having a lower velocity dispersion.

This possible difference in the properties of the emitting regions is further indicated by our RADEX models (Sect. \ref{radex}). These models reproduce the observed CH$^+$ line intensities with a temperature of $T=500-1000$ K and a density of $n\sim10^5$ cm$^{-3}$, but suggest a higher density component to explain SH$^+$ emission, $T\sim200$ K, $n\sim10^6$ cm$^{-3}$, which is consistent with the properties of warm and dense condensations suggested to explain the origin of excited OH \citep{goicoechea2011} and high-$J$ CO line emission (Joblin et al. 2012, in preparation). 
In this case, thermal line broadening may contribute to the difference between the widths of the CH$^+$ and SH$^+$ lines.
The expected contribution of thermal line broadening for CH$^+$ is $\Delta v$=$2\sqrt{2 \ln 2} \sqrt{\frac{kT}{m}}$=1.3 \kms for $T_\mathrm{kin}$=500 K, and $\Delta v$=1.8 \kms for $T_\mathrm{kin}$=1000 K. The contribution of thermal line broadening for SH$^+$ for $T_\mathrm{kin}$=200 K is $\Delta v$=0.6 \kms. 

As an alternative explanation of the large observed line width of CH$^+$, formation pumping may play a role in the broadening of CH$^+$. As explained in Sect. \ref{radex}, CH$^+$ formation results in an excess energy equivalent to 5360 K. This energy may be redistributed and go into kinetic motions. 
If the 5360 K excess energy goes into excess translational energy of the nascent CH$^+$, and if this is identified as an ionic kinetic temperature upon formation, then the corresponding FWHM of Doppler motions is 4.4 \kms.

The difference between the widths of CH$^+$ and SH$^+$ may not originate in different excitation conditions, but in the difference in the chemistry of these ions. After its formation, CH$^+$ rapidly reacts with H and H$_2$, therefore it is likely that its translational motions never become thermalized. In this case, the large velocity dispersion in CH$^+$ partially reflects the conditions of its formation. SH$^+$ on the other hand does not react rapidly with H (and H$_2$), so that it is destroyed less rapidly by recombination with electrons. Therefore, SH$^+$ can become thermalized translationally during its chemical lifetime. Therefore, while SH$^+$ traces the density and the temperature of the emitting region, CH$^+$ is more sensitive to the details of its formation process.

Another interpretation of the broadening of CH$^+$ and other molecules could be that they originate in flows created by photoevaporating clumps (e.g. \citealp{gortihollenbach2002}, \citealp{mackeylim2010}). However, our observed data don't completely support this assumption. 
If the FWHM of ions and other molecules detected at the same position had a contribution by the evaporating flow, we would expect C$^+$ and CH$^+$ to have a similar flow velocity, as the momentum transfer from H$_2$ in the H$_2^*+$C$^+$ reaction is small. Based on our HIFI observations of the CO$^+$ peak, this does not apply, as FWHM(C$^+$)$\sim$3.8 km~s$^{-1}$ and FWHM(CH$^+$)$\sim$5 km~s$^{-1}$.

The difference in the width of reactive ions tracing the warm surface region of the PDR is a key part of understanding the chemistry of these ions. Future observations of the spatial distribution of SH$^+$ will help to distinguish between these explanations.

\subsection{An extension of the `CF$^+$ ladder' in the Orion Bar}

Unlike CH$^+$ and SH$^+$, CF$^+$ does not directly form via collisions with H$_2$. The reaction between fluorine and H$_2$ is followed by a reaction between C$^+$ and HF, where HF is the dominant reservoir of fluorine and was previously been detected in emission in the Orion Bar by \citet{vandertak2012}
\[\mathrm{F + H_2 \rightarrow HF + H}\]
\[\mathrm{HF + C^+ \rightarrow CF^+ + H.}\]
CF$^+$ is the second most important fluorine reservoir and accounts for $\sim$1\% of the gas-phase fluorine abundance. However, CF$^+$ has so far only been detected in two sources. The 2-1 and 1-0 rotational transitions have been recently detected with a spatially extended emission toward the PDR in the Horsehead nebula \citep{guzman2012}. The first detection of CF$^+$ was toward the Orion Bar (\citealp{neufeld2006}), showing spatially extended emission in the 1-0, 2-1, and 3-2 rotational transitions. Our observations extend the observed CF$^+$ transitions toward the Orion Bar up to the 5-4 transition and since the line intensity is consistent with the previously observed transitions, this work gives additional confirmation on the simple CF$^+$ chemistry tracing the surface layers exposed to UV irradiation.

\section{Conclusions and outlook}

We have analyzed six rotational transitions of CH$^+$ and three transitions of SH$^+$ and have reported the first detection of the 5-4 transition of CF$^+$. Our main conclusions are the following:

\begin{itemize}

\item[-] We have detected CH$^+$ up to the 6-5 transition. The 2-1 and 1-0 transitions are spectrally resolved and show significantly broader lines ($\mathrm{\Delta v \sim 5~km s^{-1}}$) than most dense gas tracers in the Orion Bar. SH$^+$ on the other hand shows significantly narrower lines than CH$^+$ ($\mathrm{\Delta v \sim 3~km s^{-1}}$). Explanations of this difference include their origin in a different density (and temperature) component. Alternatively, because of its reactivity, CH$^+$ never becomes thermalized therefore, its observed properties trace the formation process rather than the properties of the emitting region, unlike for SH$^+$. Information on the spatial distribution of SH$^+$ is needed to resolve this puzzle.

\item[-] Inelastic collisions with H$_2$ and electrons both affect the excitation of CH$^+$ and SH$^+$, similar to the case of HF (\citealp{vandertak2012}). Reactive collisions are important in the excitation of CH$^+$, but have less effect in the case of SH$^+$.

\item[-] Comparing the observed CH$^+$ intensities to predictions of PDR models for typical conditions in the Orion Bar, we confirm that CH$^+$ forms via reactions with vibrationally excited H$_2$, as predicted by \citet{agundez2010}. Our PDR models also show that CH$^+$ forms in the warm surface region ($T\sim500-1000$ K) of the PDR at high pressures ($\sim$10$^8~{\rm{cm}}^{-3}~{\rm{K}}$).

\item[-] SH$^+$ is also likely to form via H$_2$ vibrational excitation, assuming that the formation rates are similar to that of CH$^+$. SH$^+$ is also a tracer of the warm surface regions of the PDR.

\end{itemize}

In the future, higher-resolution follow-up observations of a larger region in the Orion Bar will give more insight into the excitation conditions of SH$^+$. Probing CH$^+$ and SH$^+$ formation in PDRs with a range of parameters, such as different radiation fields would help to deepen our understanding of the chemistry of these ions in regions exposed to UV irradiation.

\begin{acknowledgements}
We thank the referee for the constructive suggestions that helped to improve the paper. We also thank the editor Malcolm Walmsley for additional comments. We thank Simon Bruderer for useful comments on CH$^+$ and SH$^+$ and Yunhee Choi for the help in the reduction of the HIFI data.
J.R.G. is supported by a Ram\'on y Cajal research contract; he thanks the Spanish MINECO for funding support through grants AYA2009-07304 and CSD2009-00038. Support for this work for E.A.B. was provided by NASA through an award issued by JPL/Caltech. Part of the work was supported by the {\it  Deutsche Forschungsgemeinschaft} through grant SFB 956 C1. \\

HIFI was designed and built by a consortium of institutes and university departments from across
Europe, Canada, and the US under the leadership of SRON Netherlands Institute for Space Research, Groningen, The Netherlands, with major contributions from Germany, France, and the US. Consortium members are Canada: CSA, U.Waterloo; France: IRAP, LAB, LERMA, IRAM; Germany: KOSMA, MPIfR, MPS; Ireland: NUIMaynooth; Italy: ASI, IFSI-INAF, Arcetri-INAF; The Netherlands: SRON, TUD; Poland: CAMK, CBK; Spain: Observatorio Astronomico Nacional (IGN), Centro de Astrobiolog\'ia (CSIC-INTA); Sweden: Chalmers University of Technology - MC2, RSS \& GARD, Onsala Space Observatory, Swedish National Space Board, Stockholm University - Stockholm Observatory; Switzerland: ETH Z\"urich, FHNW; USA: Caltech, JPL, NHSC.
HIPE is a joint development by the Herschel Science Ground Segment Consortium, consisting of ESA, the NASA Herschel Science Center, and the HIFI, PACS, and SPIRE consortia. \\

PACS was developed by a consortium of institutes led by MPE (Germany) and including UVIE (Austria); KU Leuven, CSL, IMEC (Belgium); CEA, LAM (France); MPIA (Germany); INAFIFSI/OAA/OAP/OAT, LENS, SISSA (Italy); IAC (Spain).
\end{acknowledgements}

\end{document}